\documentclass[12pt]{article}
\usepackage{makeidx}
\usepackage{multirow}
\usepackage{multicol}
\usepackage[dvipsnames,svgnames,table]{xcolor}
\usepackage{graphicx}
\usepackage{epstopdf}
\usepackage{ulem}
\usepackage{hyperref}
\usepackage{amsmath}
\usepackage{amssymb}
\usepackage[paperwidth=595pt,paperheight=842pt,top=50pt,right=70pt,bottom=70pt,left=70pt]{geometry}

\makeatletter
	{\par\setlength{\parindent}{#3}
	\setlength{\leftmargin}{#1}       \setlength{\rightmargin}{#2}%
	\advance\linewidth -\leftmargin       \advance\linewidth -\rightmargin%
	\advance\@totalleftmargin\leftmargin  \@setpar{{\@@par}}%
	\parshape 1\@totalleftmargin \linewidth\ignorespaces}{\par}%
\makeatother 

\begin{document}

\begin{center}
Why QCD Explorer stage of the LHeC should have high(est) priority
\end{center}

\begin{center}
S. A. \c{C}etin$^{a}$, S. Sultansoy$^{b}$, G. \"{U}nel$^{c}$
\end{center}

\begin{center}
$^{a}${\footnotesize  Do\u{g}u\c{s} University, Istanbul, Turkey}
\end{center}

\begin{center}
$^{b}${\footnotesize  TOBB University of Economics and Technology, Ankara, Turkey 
\\* 
and ANAS Institute of Physics, Baku, Azerbaijan }
\end{center}

\begin{center}
$^{c}${\footnotesize  University of California, Irvine, USA}
\end{center}

\textit{\noindent Abstract: The QCD Explorer will give opportunity to enlighten the origin
of the 98.5\% portion of the visible universe's mass, clarify the nature of the
strong intecartions from parton to nuclear level and provide precision pdf's for
the LHC. Especially the $\gamma{}$-nucleus option seems to be very promising for
QCD studies.}

\vspace{6pt}

\hspace{15pt}Linac-ring type colliders have two main goals: to explore TeV scale
with lepton-hadron and photon-hadron collisions and to achieve highest luminosity
at flavor factories (the history of corresponding proposals can be found in [1]).
This note concentrates on the first goal which is represented by the linac-ring
option of the Large Hadron electron Collider (LHeC), proposed to explore the
highest energy proton and ion beams available at the LHC probed by energetic
electron or gamma beams from a linac tangent to the LHC. The Conceptual Design
Report (CDR) of the LHeC project which is published in [2], investigates two
options for the collider: Linac-Ring (LR) type collider where electrons are
provided by linac and Ring-Ring (RR) option which assumes an additional electron
ring in the LHC tunnel.

\vspace{6pt}

\hspace{15pt}The idea of the LHC based linac-ring type ep/$\gamma{}$p collider
includes two stages: QCD Explorer (E$_{e }$= 60, 140 GeV) and Energy Frontier
(E$_{e }$$\geq{}$ 500 GeV). The first stage is mandatory for a deeper
understanding of the strong interactions and an adequate interpretation of the
LHC data which requires precision pdf's. The second stage, which actually depends
on the outcomes of the LHC, hence called provisional, will mainly have great
potential for BSM physics complementary to LHC and exceeding the possible ILC. It
should be noted that the Energy Frontier as well as $\gamma{}$p options of both
QCD Explorer and Energy Frontier can only be realised with the linac-ring option.

\vspace{6pt}

\hspace{15pt}Today, LR option is considered as the basic one for the LHeC.
Actually this decision was almost obvious from the beginning due to the
complications in constructing by-pass tunnels around the existing experimental
caverns and installing the e-ring in the already commisioned tunnel. Let us
remind that the CDR sthgeaof the LHC assumed also ep collisions using the already
existed LEP ring; but it turned out that LHC installation required dismantling of
LEP from the tunnel.

\vspace{6pt}

\hspace{15pt}Now that LR is the choice for the LHeC, Energy Recovery Linac (ERL)
is being pushed as a basic choice instead of the single-pass option using the
argument that it could provide an order of magnitude higher luminosity.
Nevertheless, keeping in mind that such higher luminosity is not necesseary for
the QCD Explorer, it is likely that the single-pass option will become dominant
soon, however we believe the sooner the better. It should ne mentioned that a
very important advantage of the LR option, namely the opportunity to construct
$\gamma{}$p/$\gamma{}$A collider loses its strength at rhe ERL based LHeC,
moreover the single-pass option will give the opportunity to increase the energy
of the electrons by lengthening the linac further.

\vspace{6pt}

\hspace{15pt}Concerning the physics program of the QCD Explorer, putting forward
the search for SUSY or other BSM physics or even detailed study of the Higgs
boson as the main goal would have serious drawbacks. The uniqueness of such a
machine lies in its potential to probe the nature of the strong interactions from
parton to nuclear level and provide precision pdf's for the adequate
interpretation of the LHC results. It is well known that big challenges still
exist in the QCD part of the Standard Model like understanding confinement and
quark-gluon plasma. QCD Explorer will give the opportunity to reach very small
x$_{g}$ region [3] shedding light on confinement. Then according to vector meson
dominance the $\gamma{}$A collider will act as a $\rho{}$A collider which will
give an opportunity to investigate formation of the quark gluon plasma at very
high temperatures and low densities.

\vspace{6pt}

\hspace{15pt}In light of the discussions presented above we propose the
following phases for QCD Explorer based on single-pass linac option. First phase:
ep collider with luminosity of 10$^{32}$cm$^{-2}$s$^{-1}$ and eA collider with
luminosity of AxL$_{eA}$=10$^{31}$cm$^{-2}$s$^{-1}$ which seems sufficient for
QCD studies. Second Phase: $\gamma{}$p and $\gamma{}$A collider with similar luminosities.
Third Phase: construction of a second single-pass linac for energy recovery [4]
to achieve much higher luminosities. Fourth Phase: lengthening the single-pass
linac to switch to Energy Frontier stage.

\vspace{6pt}

\hspace{15pt}With the discovery of the long sought Higgs boson, the electroweak
sector of the Standard Model has filled its gaps. At this point it is worth
mentioning that the Higgs Mechanism accounts for only $\sim$1.5\% of
the mass of the visible universe and the rest, $\sim$98.5\% is
provided by the QCD. Hence another strong motivation of the QCD Explorer is to
better understand the formation of the visible universe.

\vspace{6pt}

\hspace{15pt}In conclusion, we hope that the presented qualitative arguments
justify the necessity of the QCD Explorer for the future of the high energy
physics.
\vspace{12pt}

\textbf{References}

\vspace{6pt}

1. A.N. Akay, H. Karadeniz and S. Sultansoy, \textit{Review of Linac-Ring Type
Collider Proposals}, Int. J. Mod. Phys. A25 (2010) 4589-4602; e-Print:
arXiv:0911.3314 [physics.acc-ph]

2. J L Abelleira Fernandez et al. (LHeC Study Group), \textit{A Large Hadron
Electron Collider at CERN: Report on the Physics and Desigon Concepts for Machine
and Detector}, J. Phys. G. 39 (2012) 075001; e-Print: arXiv:1206.2913
[physics.acc-ph]

3. U. Kaya, S. Sultansoy, G. Unel,  \textit{Probing small x(g) region with the
LHeC based gamma-p colliders}, Nov 2012, e-Print: arXiv:1211.5061 [hep-ph]

4. V. Litvinenko, \textit{LHeC with \textasciitilde{}100\% energy recovery
linac}, 2nd CERN-ECFA-NuPECC workshop on LHeC, Divonne-les-Bains, 1-3 Sep (2009).

\end{document}